\documentclass[aps,prl,showpacs,twocolumn,groupedaddress,amssymb]{revtex4}
\usepackage{graphicx}% Include figure files
\usepackage{dcolumn}% Align table columns on decimal point
\usepackage{bm}% bold math

\begin{document}

\title{THz radiation from two-colored lasers: Non-linear convection}% \\

\author{S. Son}
\affiliation{169 Snowden Lane, Princeton, NJ, 08540}
%\author{Sung Joon Moon}
%\affiliation{PACM, Princeton University, Princeton, NJ 08544}
%\author{J.~Y. Park}
%\affiliation{Los Alamos National Laboratory}

\begin{abstract}
The non-linear convection for the THz light generation from  two-colored lasers are analyzed in the context of  the forward Raman scattering. 
The energy transfer from the lasers to the THz light can be efficient. The possible peak intensity of the generated THz light is estimated and the optimal duration time is estimated.   
%A new scheme for soft x-ray lasers  is proposed. The backward Haman scattering between an intense visible-light laser and a relativistic electron beam results in soft x-ray light via the Doppler shift. One of the most intense soft x-ray light sources is contemplated.  
\end{abstract}

\pacs{42.55.Vc, 42.65.Dr,42.65.Ky, 52.38.-r, 52.35.Hr}       

\maketitle

%There are many potential applications with intense THz light~\cite{siegel,siegel2,  siegel3, booske,radar, diagnostic, security} and great advances for the light source have been made by two-color lasers;
%After injecting $\omega$ and $2 \omega$ lasers into a plasma, 
Very bright
THz light is observed in  two-color lasers, for which 
the transient photo-ionized current is responsible~\cite{kim}.
% where the conventional linear analysis fails to account for~\cite{kim2}.
In this paper, the author provides another intense  THz light generation by two-colored lasers in a fully ionized plasma. 
%Two colored lasers co-propagating beat each other and emit a THz light in the same direction. 
% The author consider the THz light co-propagating with the lasers. 
The analysis in this paper shows that  the emission due  to the forward Raman scattering could be very strong; 
% intenisty (the interaction) is intense (long) enough, an intense THz light can be generated. 
%It will be shown that the THz light generation in this context can be described by the physics similar to the forward Raman scattering (FRS); 
Using the non-linear theory of the FRS   
%It is shown that the peak inetnsity of the THz light achieved can be as intense as the photo-ionization current case.
 the  peak intensity achievable  is theoretically estimated. 
%  For an appropriate physical parameter relevant to the real experiments, 
%10 percents of the theoretical peak is shown to be  achieved, illustrated by 1-d simulation. 
% The reason that the  THz light intensity can not reach the theoretical limit is due to the 
The author identifies  a non-linear convection which  suppresses the energy channeling, thus limiting the peak THz intensity.  
An alternative method, by which this non-linear convection is suppressed and the intensity peak can reach the theoretical limit, is proposed and validated by 1-d simulation.  
%While there are some non-linear convection mechanism, reducing the energy conversion from the lasers to the THz, 
%an alternative method, by which the theoretical limit can be reached, is proposed  validated by 1-d simulation.  
Very high  intensity can be reached, and  the duration (energy) of the pulse could be much longer (higher) than the one achieved by the photo-ionization case.  The current scheme  can be also used for the generation of the far infra-red light.

Consider a familiar situation where 
 two colored lasers with the frequency $\omega_1 $ and $\omega_2$ co-propagates.  Their frequencies are such that   
  $|2 \omega_1 - \omega_2| \ll \omega_1$  and $|2 \omega_1 - \omega_2| \ll \omega_2$. 
They beat each other producing a THz light with the frequency $\omega_3 =|2 \omega_1 - \omega_2|$ by various mechanisms such as the linear beating or the photo-ionization. 
The author claims that the Raman scattering could be an important physics to consider in the two-color laser THz light generation:  the plasma density
perturbation,  via the interaction between the lasers, 
and a THz current by the density perturbation and  laser quivers.
%  the two lasers excite the plasma density via the beating ponderomotive potential, which produces a THz current with one laser quiver.
In this paper, the author considers the situation when  the THz light  co-propagates and interacts with the lasers via the FRS. 
If the lasers are intense enough and coherent enough,  this THz  emission by the lasers could be stronger than the broad-band emission.  The author analyzes the efficiency and drawback of the emission processes. 
%Consider two colored lasers co-propagating beat each other and emit a THz light in the same direction. 
%Furthermore, the intensity of the lasers could be very high so that the quiver velocity is comparable to the velocity of light. Then, the THz intensity becomes very high even if the interaction time is short.  

The non-linear theory of the laser interaction in plasmas have been developed in the laser-plasma physics community for a few decades because of the relevance to the inertial confinement fusion~\cite{tabak}, including  the backward (forward) Raman (Brillouin) scattering~\cite{Kruer,sonforward, sonbackward, McKinstrie,malkin1}. 
When everything is co-propagating, the forward Raman scattering (FRS) is the appropriate physics. 
%The FRS, as one of the dominant non-linear interaction between intense lasers,  is well-established and thoroughly studied for a few decades. 
In the FRS, two lasers excite the plasma wave via the beating ponderomotive potential. This plasma wave generates a beat current with one laser quiver,  transferring energy to the other laser~\cite{McKinstrie}.
The main differences  between the FRS and the current situation is that 
1) the density perturbation here is not a resonant plasma wave and   
% However, the FRS physics still can be used if the plasma wave is replaced by the density perturbation. 
2) there are three E\&M fields in the current case while there is only two E\&M field in the FRS.
There are three E\&M waves and three density perturbations from ponderomotive interactions of the E\&M waves. 
In total, there are  9 possible combinations of the wave interactions and  
all those non-linear convection terms  needs to be properly taken into account in order to analyze the physics and efficiency of the THz light generation. 
%  There are many  non-linear interaction terms. 

For simplicity, let us assume that the lasers are linearly polarized and  their electric fields are parallel to each other.  The density response to the ponderomotive potential can be obtained from the continuity equation and the momentum equation: 

\begin{eqnarray}
 \frac{\partial \delta n_e }{\partial t}  &=&-  \mathbf{\nabla} \cdot ( \delta n_e \mathbf{\mathrm{v}} ) \nonumber \mathrm{,}\\ \nonumber \\
 m_e \frac{d \mathbf{v} }{dt} &=& e\left( \mathbf{\nabla} \phi  - \frac{\mathbf{v}}{c} \times \mathbf{B}\right) \nonumber \mathrm{,} \\ \nonumber
\end{eqnarray}
Combining the above equations with the Poisson equation $\nabla^2 \phi = -4 \pi \delta n_e e $, the density response can be obtained from

\begin{equation} 
\left(\frac{\partial^2 }{ \partial t^2 } + \omega_{\mathrm{pe}}^2\right)\delta n_e =
\frac{en_0 }{m_ec} \mathbf{\nabla} \cdot \left( \mathbf{v} \times \mathbf{B} \right)\mathrm{.} \label{eq:den}
\end{equation} 
where $\omega_{\mathrm{pe}}^2= 4\pi n_0 e^2/ m_e$ is the plasma Langmuir wave frequency and $n_0$ is the background electron density. 
The density perturbation from the ponderomotive force of  any one pair of the lasers ($1,2,3$) can be  estimated from Eq.~(\ref{eq:den}). For an example,  the density perturbation from $1,2$ with the frequency $\omega_n = \omega_1 + \omega_2 $ is given as~\cite{McKinstrie}

\begin{eqnarray} 
 \delta n(\omega_1 + \omega_2, k_1 + k_2)  &=& -n_0 \frac{\omega_n^2 }{ \omega_n^2 - \omega_{\mathrm{pe}}^2} a_1 a_2  \nonumber \\ \nonumber \\
&\cong& -n_0 a_1 a_2 \mathrm{,} \label{eq:den2}
\end{eqnarray} 
where $a_{1,2} = eE_i/m\omega_{1,2} c $ is the laser quiver velocity normalized by the velocity of the light, $E_{i}$ is the electric field of the laser $i$,  and $\delta n \cong -n_0 a_1 a_2 $ because $\omega_{1,2,3} \gg \omega_{\mathrm{pe}}$.  Similarly,  the density perturbation for $\omega_n = \omega_2-\omega_1$ is given as $ \delta n = -n_0 (\omega_n^2/ (\omega_n^2 -\omega_{\mathrm{pe}}^2)) a_1^{*} a_2 \cong -n_0     a_1^{*} a_2$, where  $a_1^{*}$ is the complex conjugate.

Using Eq.~(\ref{eq:den2}) and   the FRS theory, the non-linear convection equation can be derived by taking out the fast time scale~\cite{McKinstrie};  
\begin{eqnarray}
L_1 a_1  &=& +i \frac{\omega_{\mathrm{pe}}^2 }{2\omega_1} \left(2 (|a_2|^2 +  |a_3|^2) a_1  +  3  a_2 a_3  a_1^{*}\right)    \label{eq:1} \mathrm{,}\\ \nonumber \\ 
L_2 a_2  &=& +i \frac{\omega_{\mathrm{pe}}^2 }{2\omega_2}  \left(2 (|a_1|^2 + |a_3|^2) a_2 +  2 a_1^2 a_3^{*}\right)     \label{eq:2} \mathrm{,}\\ \nonumber \\ 
L_3 a_3  &=& +i \frac{\omega_{\mathrm{pe}}^2 }{2\omega_3}  \left(2(|a_1|^2 +|a_2|^2) a_3  + 2 a_1^2 a_2^{*}\right)   \label{eq:3}  \mathrm{,}\\ \nonumber 
\end{eqnarray}
where $L_{i} = (\partial/\partial t) + v_{i} (\partial/\partial z) $ with $v_{i}$ is the group velocity of the laser.  In the above equations, the decay of the lasers, such as the inverse bremsstrahlung or the non-linear decays, is assumed to be small.

Some terms in Eqs.~(\ref{eq:1}), (\ref{eq:2}) and  (\ref{eq:3}) need more explanation. The first and the second  terms on the right side of  Eqs.~(\ref{eq:1}), (\ref{eq:2}) and  (\ref{eq:3})  neither   change amplitudes  ($|a_i|^2$) nor  transfer any energy between different lasers. They only  change the phase of each laser.  These  are the original FRS terms between one pair of lasers;  a density perturbation is caused by the ponderomotive force between  one pair (among 1,2,3) and the beating of the density perturbation with the quiver velocity of one of the pair modifies  the phase of the other.
However, unlike the conventional FRS, these terms do not transfer energy between lasers because  a density perturbation  is not a resonant plasma wave.

More interesting ones are  the third term on the right side of Eqs.~(\ref{eq:1}), (\ref{eq:2}) and  (\ref{eq:3}).
They represents actual non-linear channeling between the lasers and THz light. 
They stem from the same physics with the first and the second term; the ponderomotive force of the lasers or THz light creates the density perturbation by Eq.~(\ref{eq:den}) and the beating of the density perturbation with the quiver transfers the energy from one to another. The only difference between the first (the second) term and the third term is the fact that
the first (second) term only involves one pair of lasers, while the third them involve all three E\&M waves.  
For an example, consider the third terms of Eq.~(\ref{eq:1}), $a_2a_3 a_1^*$. This term represents three possible beating interactions;
the beat current between a density perturbation $\delta n$, with the frequency  ($\omega_2+\omega_3$, $\omega_1-\omega_3$, $\omega_1+\omega_2$),  and  the laser quiver ($\omega_1$, $\omega_2$, $\omega_3$). The coefficient $3$ stems from these three possible interactions.  The other terms can be also explained in a similar way.  The above equations are exact to the third order.

\begin{figure}
\scalebox{0.45}{
\includegraphics[width=1.7\columnwidth, angle=270]{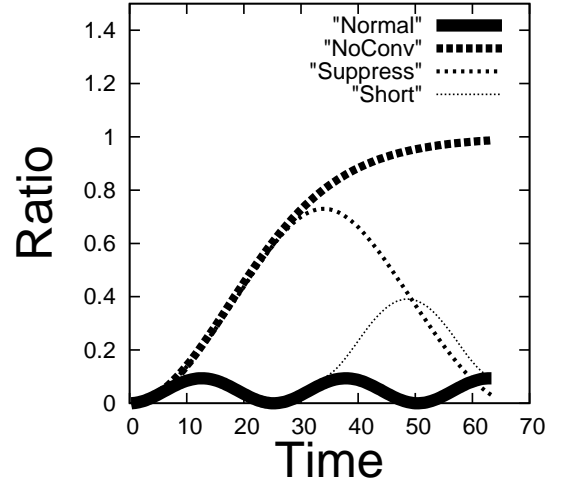}}
\caption{\label{fig:1}
2.5 THz light emission
}
\end{figure}

The first obvious observation of Eqs.~(\ref{eq:1}), (\ref{eq:2}) and  (\ref{eq:3}) is that there are local conserved quantities.
$N_1 = \omega_1 |a_1|^2 + 1.5 \omega_2|a_2|^2$ ($ N_2 = \omega_1 |a_1|^2 +1.5 \omega_3|a_3|^2$). Especially, the following relation can be deduced: 

\begin{equation} 
\omega_3 |a_3(t)|^2 =  \frac{2}{3} \left( \omega_1 |a_1(t=0)|^2 - \omega_1 |a_1(t)|^2 \right) \mathrm{.}
\label{eq:4}
%I_3  = \omega_2 |a_2|^2 -\omega_3 |a_3|^2 \label{eq:con} \mathrm{.}
\end{equation}
From Eq.~(\ref{eq:4}),  the theoretical possible intensity peak of the THz light is given as 

\begin{equation}
  |a_T|^2  = \frac{2\omega_1}{3\omega_3}  |a_1(t=0)|^2 \mathrm{.} \label{eq:5}
\end{equation}

An illustrative example of the solution for Eqs.~(\ref{eq:1}), (\ref{eq:2}) and  (\ref{eq:3}) is provided in Fig.~(\ref{fig:1}).  Assuming a homogeneous E\&M wave, the ordinary equation of three complex variables can be  solved. 
  In this example, the  $\omega_1 $ laser has the wave length of $7.74 \ \mathrm{\mu m}$  with the intensity of $I = 5 \times 10^{15} \  \mathrm{W}  / \mathrm{cm}^2 $ and    the  $\omega_2 $ laser   has the wave length of $4.0 \  \mathrm{\mu m}$ with the intensity of $I =  10^{16} \ \mathrm{W}  / \mathrm{cm}^2 $. Given the laser intensity and the frequency,  $a_1 \cong 0.5 $ $a_2 \cong 0.4$.  The generated THz light has the frequency ($2 \omega_1 -\omega_2 $) of 2.5 THz. 
The  plasma density  is assumed to be  $n_0 \cong 3 \times  10^{15} \  / \mathrm{cm}^3 $. 
Given these information, the ordinary differential equation is  solved for $|a_3|^2$.  We plot $|a_3|^2 /   |a_T|^2$ as a function of time in Fig.~(\ref{fig:1}). The time is normalized by pico-second in Fig.~(\ref{fig:1}) and it is labeled as ``Normal''.   
As shown, the THz light intensity oscillates and attain, in maximum, 10 percents of the theoretical peak limit given in Eq.~(\ref{eq:5}).
The peak intensity attained (theoretical peak limit) is  $I \cong 2\times  10^{13} \  \mathrm{W}  / \mathrm{cm}^2 $ ($I \cong 2\times  10^{14} \  \mathrm{W}  / \mathrm{cm}^2 $).  
 The time duration for which the peak can be achieved is 10 pico-seconds. 
The length of the plasma for such duration is roughly  0.5 mm ($L = c \delta t$). 
%The peak intensity corresponds to $I \cong 2\times  10^{13} \  \mathrm{W}  / \mathrm{cm}^2 $. 

In the above example, the laser intensity is high so that the the interaction time is short and the necessary size of the plasma is small. 
However, given the fact that, as  three waves are co-propagating, the coherence will be preserved in time, even moderately intense laser would efficiently channel their energy into the THz light. 
In other words, 
 from  Eqs.~(\ref{eq:1}), (\ref{eq:2}) and  (\ref{eq:3}), 
 if $a_1(t)$, $a_2(t)$ and $a_3(t)$ is a solution, 
 $\eta a_1(\eta^2t)$, $\eta a_2(\eta^2t)$ and $\eta a_3(\eta^2 t)$ is also a solution. The convection time, in which the THz light attain its peak, can be roughly estimated as $\delta t \cong (\omega_{\mathrm{pe}^2}/ \omega_1 \omega_3)(1/|a_1(t=0)|^2)$, which is inversely proportional to $|a_1(t=0)|^2$. In the example of Fig.~(\ref{fig:1}), if the intensity is reduced by 100, the necessary size of the plasma would be 60 cm.

By experimenting with Eqs.~(\ref{eq:1}), (\ref{eq:2}) and  (\ref{eq:3}), a few useful conclusions about the optimal parameterization can be drawn. 
First, the optimal range for the $a_1(t=0)$ and $a_2(t=0)$ are 
$0.5 |a_1| < |a_2| < |a_1| $ with a view to maximizing the peak THz light normalized by the theoretical peak limit in Eq.~(\ref{eq:5}). Given the fact that $2\omega_1\cong \omega_2$, the optimal operating regime is when the $\omega_2$-laser intensity is one or two times higher than the $\omega_1$-laser.  
Another useful conclusion is that  that the non-linear convection terms, the first and the second terms of the right hand side in Eqs.~(\ref{eq:1}), (\ref{eq:2}) and  (\ref{eq:3}), keeps the THz from attaining the theoretical peak limit. 
If those terms are taken out and the equations are solved, the 100 percents of the theoretical limit can be reached, labeled as ``NOConv'' in Fig.~(\ref{fig:1}).  Especially, by just taking out the convection terms in Eq.~(\ref{eq:3}), 80 percents of the theoretical limit can be reached.  Then, it would be beneficial if we suppress those terms in a certain way (especially those in equation~\ref{eq:3}).  One such method is presented in the below. In particular, the authors tries to suppress  the convection terms in Eq.~(\ref{eq:3}).

\begin{figure}
\scalebox{0.45}{
\includegraphics[width=1.7\columnwidth, angle=270]{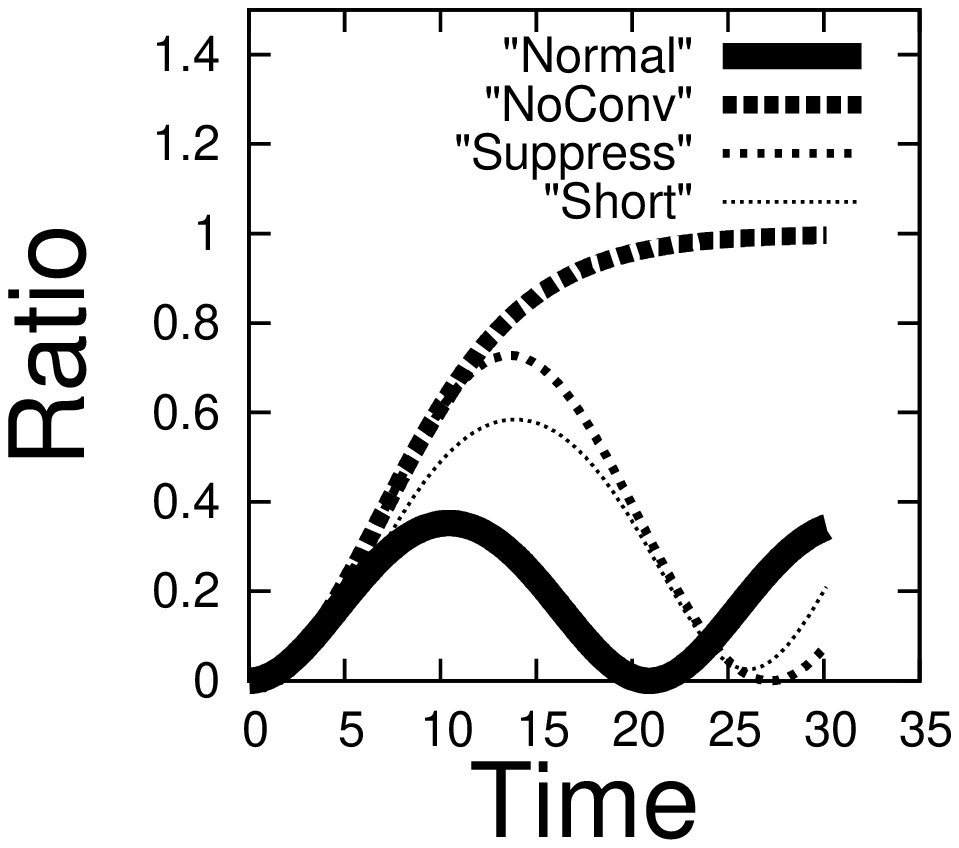}}
\caption{\label{fig:2}  33 $\mathrm{\mu m}$ Far infra-red emission
}
\end{figure}

It can be deduced from Eq.~(\ref{eq:den2}) that 
the density perturbation $\delta n$ could have the opposite sign from the first and the second term if $\omega_n < \omega_{\mathrm{pe}} $.  So, consider a situation where another moderately intense THz light with $\omega_4$ is counter-propagating (co-propagating) with the lasers.  Then, Eq.~(\ref{eq:3}) is modified to 

\begin{equation}
L_3 a_3  = +i \frac{\omega_{\mathrm{pe}}^2 }{2\omega_3}  \left((|a_1|^2 +|a_2|^2-\alpha^2) a_3  + 2 a_1^2 a_2^{*}\right)   \label{eq:33} \mathrm{,}
\end{equation}
where $\alpha^2 = \left(\omega_N^2 / (\omega_{\mathrm{pe}}^2 - \omega_n^2)\right)|a_4|^2 $, 
$\omega_n = |\omega_3 -\omega_4 |$ and  $  \omega_N = \omega_3 +\omega_4$ ($\omega_N =\omega_n $) for counter-propagating (co-propagating) THz light.  
%\bibliography{tera2}% Produces the bibliography via BibTeX.

The idea for the suppression is to render $|a_1|^2 +|a_2|^2-\alpha^2$ small on average.
The various 1-d simulation suggest that the optimal value of $\alpha^2$ is 
such that  $
0.5 (|a_1(t=0)|^2 +|a_2(t=0)|^2)<   \alpha^2 < (|a_1(t=0)|^2 +|a_2(t=0)|^2) $
  One example when  $\alpha^2= 0.7 (|a_1(t=0)|^2 +|a_2(t=0)|^2)$ is shown in Fig~(\ref{fig:1}) labeled as ``Suppress'', in which the 80 percents of the theoretical peak is achieved. The time it took for the peak THz light is 30 pico-second, which corresponds to the plasma size of 0.6 cm. 
The above condition  requires the intensity of the THz light $\omega_4$ to be a certain range.  Assuming $\delta \omega = \omega_{\mathrm{pe}} - |\omega_3 - \omega_4| \ll \omega_3$, 
it can be re-casted as $0.5 (\omega_{\mathrm{pe}} \delta \omega/
\omega_N^2 )(|a_1(t=0)|^2 +|a_2(t=0)|^2)  < |a_4|^2 < 
(\omega_{\mathrm{pe}} \delta \omega/ \omega_N^2 ) (|a_1(t=0)|^2 +|a_2(t=0)|^2)$. From Eq.~(\ref{eq:5}), assuming$|a_1(t=0)|^2 \cong |a_2(t=0)|^2 $, this condition is the same as

\begin{equation}
|a_4|^2 \cong \left(\frac{\omega_3}{\omega_1} \right) \left( \frac{\omega_{\mathrm{pe}} \delta \omega }{\omega_N^2}  \right) |a_T|^2 \mathrm{.} \label{eq:cond} 
\end{equation}
For the example in Fig.~(\ref{fig:1}), $\omega_3 /\omega_1 \cong 1/15$ and $\omega_{\mathrm{pe} } / \omega_3 \cong 1/5 $. 
If $\delta \omega \cong 0.1 \omega_{\mathrm{pe} }$, 
then optimal intensity of $\omega_4$ THz light  is given as  $|a_4|^2 \cong 0.00003 |a_T|^2$ ($|a_4|^2 \cong 0.003 |a_T|^2$) for counter-propagating (co-propagating) one; 
a low-intense THz light counter-propagating (co-propagating) can suppress 
the non-linear convection for enabling the THz light reach  the theoretical peak limit.

As shown in Eq.~(\ref{eq:cond}), the intensity requirement for counter propagating THz light is lower than the co-propagating beam by a factor of $4 (\omega_4/\omega_{\mathrm{pe}})^2$. However, the duration of the counter propagating beam seems to be necessarily long in order to suppress the non-linear convection during the entire interaction time. This turns out not to be the case. 
In Fig.~(\ref{fig:1}),  we solve a case of a counter-propagating THz light with $ \alpha^2 \cong 3.0  (|a_1(t=0)|^2 +|a_2(t=0)|^2)$ for a brief portion of the interaction; more specifically between 25 pico-second and 28 pico-second, labeled as ``Short''.  In this case, 40 percents of the theoretical peak is achieved. 
This means that extremely low intensity counter propagating THz light could suppress the non-linear convection efficiently at  appropriate timing.
The counter-propagating beam might be more advantageous  due to the low intensity requirement.
One more interesting aspect of ``Short'' is that the THz light intensity is staying above the 10 percents of the theoretical peak after the interaction with the $\omega_4$ field; some ''hysteresis effect might be in play.  
The system of  Eqs.~(\ref{eq:1}), (\ref{eq:2}) and  (\ref{eq:3}) seems to be interesting as a dynamical system and the analytical characterization is the beyond the scope of this paper.

In Fig~(\ref{fig:2}), the far infra-red emission is studied. 
In this example, the  $\omega_1 $ laser has the wave length of $7.05 \ \mathrm{\mu m}$  with the intensity of $I = 4 \times 10^{15} \  \mathrm{W}  / \mathrm{cm}^2 $ and    the  $\omega_2 $ laser   has the wave length of $4.0 \ \mathrm{\mu m}$ with the intensity of $I =  10^{16} \ \mathrm{W}  / \mathrm{cm}^2 $
so that   $a_1(t=0) \cong 0.5 $ $a_2(t=0) \cong 0.4$.   The generated infra-red light  has the wave frequency (length) of  10 THz (30 $\mathrm{\mu m}$).
A plasma density of $n_0 \cong 5 \times  10^{16} \  / \mathrm{cm}^3 $ is considered. 
% The generated infra-red light  has the wave length of  30 $\mathrm{\mu m}$. 
We plot $|a_3|^2 /   |a_T|^2$ as a function of time (``Normal''). The time is normalized by pico-second.   
The peak intensity attained (theoretical limit) is  $I \cong 3 \times  10^{14} \  \mathrm{W}  / \mathrm{cm}^2 $ ($I \cong  10^{15} \  \mathrm{W}  / \mathrm{cm}^2 $).  
 The time duration for which the peak can be achieved is 7 pico-seconds. The length of a plasma for such duration is roughly  2 mm. 
%The peak intensity corresponds to $I \cong 2\times  10^{13} \  \mathrm{W}  / \mathrm{cm}^2 $. 
  One example when  $\alpha^2= 0.7 (|a_1(t=0)|^2 +|a_2(t=0)|^2)$ is shown in Fig~(\ref{fig:2}) labeled as ``Suppress'', in which the 80 percents of the theoretical peak is achieved. The time it took for the peak THz light is 12 pico-second, which corresponds to the plasma size of 0.35 cm. 
A counter-propagating THz light with $ \alpha^2 \cong 3.0  (|a_1(t=0)|^2 +|a_2(t=0)|^2)$ for a brief portion of the interaction is also presented; more specifically between 8 pico-second and 9 pico-second, labeled as ``Short''.  In this case, 80 percents of the theoretical peak is achieved.
The Fig.~(\ref{fig:2}) suggests that the current scheme works well in the far infra-red emission. Notable differences from the THz emission is 1) the time it takes to attain the peak is shorter  and 2) the attained peak (``Normal'') is much higher. As the frequency of the infra-red is higher, the coupling between the emitted light with the lasers becomes stronger. 

In order to understand  why the full non-linear theory is necessary, 
 the limitations of the conventional linear theory  needs to be accounted for the current scenario.
% with a view to  understanding  why the full non-linear tehory is necessary.
%The intensity of the lasers could be very high so that the quiver velocity is comparable to the velocity of light. Then, the generated THz intensity becomes very high even if the interaction time is short. The non-linear convection cannot be ignored anymore.   
The lasers and the THz light generated are co-propagating with the same speed, the THz light strength grows linearly as the co-propagating lasers amplify coherently. If the interaction time is long enough, the THz light intense will get intense enough to feedback the lasers.  The conventional linear theory predicts the linearly growing THz light without limit, but the growing without limit becomes quickly invalid when all waves are co-propagating and lasers are intense.

In this paper, THz light generation in the two-colored lasers is proposed based on the physics similar to the forward Raman scattering and  
the non-linear convection of the energy between the lasers and the THz light is analyzed. The theoretical limit of the achievable THz light peak has been estimated and it is shown that 10 percents of the limit can be achieved in a appropriate plasma density, and laser intensities. 
The non-linear saturation, which suppresses the THz light intensity, is identified and a method to mitigate the non-linear saturation, is proposed by an additional low-intensity THz light in the counter-propagating (co-propagating) direction. In the current scenario, the achievable peak intensity is very high, roughly a few percents  or 10 percents of the laser intensity.

There are three key advantages of the current scheme over the photo-ionization case. First,  since the interaction required is non-transient, the pulse duration of the generated THz light can be much longer and thus the energy that could be carried by THz light could be much higher. 
Second, the intensity of the laser could be lower than the photo-ionization threshold if the lasers interact in a sufficiently large plasma. 
Third, as shown in Fig.~(\ref{fig:2}), the current scheme can be used for the generation of the coherent  far infra-red light. 
There could be  many potential applications with the current scheme~\cite{siegel,siegel2,  siegel3, booske,radar, diagnostic, security}.  

The author would like to point out the limitation of the current analysis in this paper.  The non-linear convections treated here are the third order of the visible light laser (THz light laser) intensity. Even though  all physics are treated as the classical physics in this paper, there are many third order terms ignored due  to the relativistic effect on  the ponderomotive potential and the electron movement in the presence of the E\&M field.  The proper theoretical treatment of these is non-trial and needs to be worked out in the future, which is beyond of this paper.

\bibliography{tera2}% Produces the bibliography via BibTeX.

\end{document}